\documentclass[conference]{IEEEtran}
\IEEEoverridecommandlockouts
\usepackage{cite}
\usepackage{amsmath,amssymb,amsfonts}
\usepackage{algorithmic}
\usepackage{graphicx}
\usepackage{textcomp}
\usepackage{xcolor}
\usepackage{multirow}
\usepackage{caption}
\usepackage[font=small,labelfont=bf]{caption}
\usepackage{subcaption}
\def\BibTeX{{\rm B\kern-.05em{\sc i\kern-.025em b}\kern-.08em
    T\kern-.1667em\lower.7ex\hbox{E}\kern-.125emX}}
\begin{document}
\title{Wireless Mobile Distributed-MIMO for 6G 
}

\author{\IEEEauthorblockN{
         Kumar Sai Bondada\IEEEauthorrefmark{1},
         Daniel J. Jakubisin\IEEEauthorrefmark{1}\IEEEauthorrefmark{2},
         Karim Said\IEEEauthorrefmark{1},
         R. Michael Buehrer\IEEEauthorrefmark{1},
         Lingjia Liu\IEEEauthorrefmark{1}     
} 
     \IEEEauthorblockA{
         \IEEEauthorrefmark{1}Wireless@VT, Bradley Department of ECE, Virginia Tech, Blacksburg, VA, USA \\
         \IEEEauthorrefmark{2}Virginia Tech National Security Institute, Blacksburg, VA, USA  
         } 
         }
\vspace{-5in}
\maketitle
\begin{abstract}

The paper proposes a new architecture for Distributed MIMO (D-MIMO) in which the base station (BS) jointly transmits with wireless mobile nodes to serve users (UEs) within a cell for 6G communication systems. The novelty of the architecture lies in the wireless mobile nodes participating in joint D-MIMO transmission with the BS (referred to as D-MIMO nodes), which are themselves users on the network. The D-MIMO nodes establish wireless connections with the BS, are generally near the BS, and ideally benefit from higher SNR links and better connections with edge-located UEs. These D-MIMO nodes can be existing handset UEs, Unmanned Aerial Vehicles (UAVs), or Vehicular UEs. Since the D-MIMO nodes are users sharing the access channel, the proposed architecture operates in two phases. First, the BS communicates with the D-MIMO nodes to forward data for the joint transmission, and then the BS and D-MIMO nodes jointly serve the UEs through coherent D-MIMO operation. Capacity analysis of this architecture is studied based on realistic 3GPP channel models, and the paper demonstrates that despite the two-phase operation, the proposed architecture enhances the system’s capacity compared to the baseline where the BS communicates directly with the UEs.

\end{abstract}

\section{Introduction and Related Works}

In Distributed MIMO (D-MIMO) systems, a group of distributed antennas using different radios is spread out in a geographic area to cooperate and form a large virtual antenna array, aiming to achieve capacity and reliability gains promised by MIMO while circumventing the form factor constraints of individual base stations (BSs). In the typical formulation, these radio units (RUs) are connected to a single central signal processing unit (CPU) through a high-speed front-haul (FH), utilizing dedicated, expensive coaxial or fibre optic cables. Research \cite{6495775,5594708,5490977, 4557195,1207650,4202181} has shown promising gains from an information-theoretic point of view, as well as improvements in coverage and range extension with uniform quality of service (QoS). However, the challenges \cite{6804225,6574665} of this system include increased complexity and significant signalling overhead (due to a greater number of RF chains) and necessitating extensive infrastructure deployments (requiring physical space for antenna deployment with the high-capacity FH links, including the cost of real estate rental and high installation time). This demands substantial network optimization to determine the placement of these static radios, resulting in escalated CapEx and OpEx costs. Alternatively, wireless D-MIMO was proposed with multiple RUs distributed over an area connected and controlled by a CPU through wireless FH links. In prior literature, the CPU transmits data to RUs and a group of RUs then serve the users (UEs). The RUs are referred to as Access Points (APs) and the UEs define which APs to serve them. This approach was explored in user-centric cell-free Massive MIMO with immobile APs \cite{9786576,10066319,9615200} and with mobile APs using Unmanned Aerial Vehicles (UAVs) \cite{8660516,9880734,10366311,8683875}.

\begin{figure*}[h]
    \centering
    \includegraphics[width=0.9\textwidth]{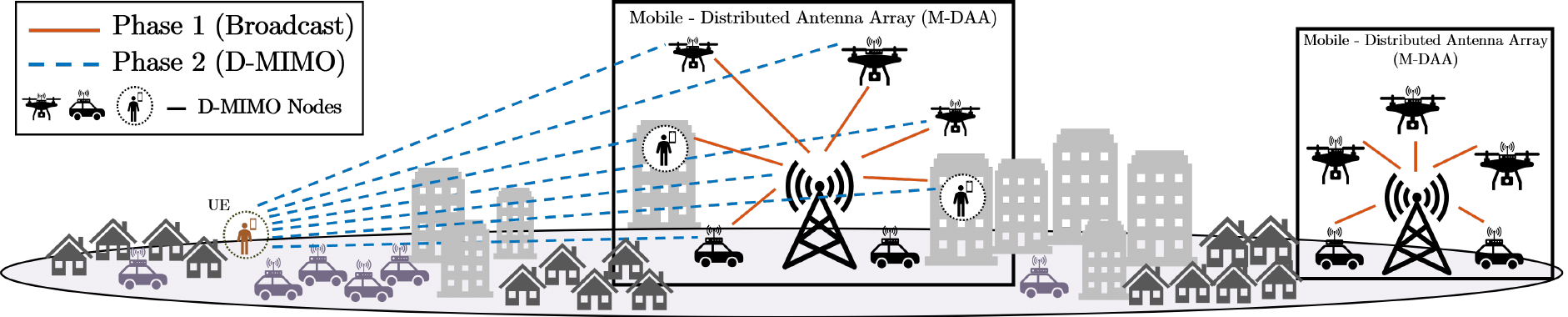}
    \caption{Wireless D-MIMO architecture showing two phases, Phase 1 and Phase 2, the mobile D-MIMO nodes cooperating with the Base station forming Mobile-Distributed Antenna Array (M-DAA)}
    \label{fig:vtc_system_model}
        \vspace{-0.2in}
\end{figure*}

Both wired and wireless D-MIMO systems require substantial modifications to current network architectures by distributing APs throughout an entire cell and connecting them to the CPU via wired or wireless FH links. We propose an architecture where the users on the network are wireless mobile nodes (D-MIMO nodes, which act as transceivers with multiple antennas) that participate in joint distributed MIMO transmission with the BSs. Our architecture differs from previously proposed D-MIMO systems, as we consider the BS to take on the responsibilities of the CPU while also jointly transmitting with the D-MIMO nodes. In other words, the RU on the BS and the RUs of the D-MIMO nodes jointly transmit together, with precoding occurring at each entity. Our architecture operates in two phases: an initial communication phase between the BS and D-MIMO nodes to forward data for joint transmission, i.e., front-haul, followed by a D-MIMO phase where the BS and D-MIMO nodes jointly transmit to the UEs, i.e., access. Unlike existing D-MIMO work \cite{9786576,9048836,10066319}, we assume that both phases share the same frequency band since the D-MIMO nodes are users on the network. Thus, the initial transmission phase is a cost (loss of channel resources) that must be accounted for in the capacity analysis. To minimize this cost, the D-MIMO nodes are selected such that the initial phase is a high-capacity link between the BS and D-MIMO nodes, i.e., closer to the BS. Capacity analysis considering both phases is studied in our paper and is compared with a baseline where the BS communicates directly with the UEs without the D-MIMO nodes.

We consider utilizing network-assisted D-MIMO nodes, which can connect to UEs in the network. These nodes may include 3GPP-oriented UAVs \cite{3gpp.22.125}, Vehicular UEs \cite{9392777}, and existing handset UEs. The nodes can be either opportunistically selected or dedicated. For instance, vehicular or handset UEs can be opportunistically chosen, or the BS can control drones for optimized placement. Moreover, the D-MIMO nodes are not restricted to a single BS, they can be capable of functioning with any BS. The D-MIMO nodes can be deployed on-demand in areas where the network load on the BS is high and the BS alone is insufficient to support UEs in the geographic area or support the users who are in a complete blackout region due to low SNR links from the BS. Since the nodes are wirelessly connected to the BS, this reduces infrastructure costs associated with deploying fiber networks, provides the required deployment flexibility with optimized placement, and offers the scalability to increase or decrease the number of nodes as needed to scale up or down the antennas. The architecture incurs signalling overhead; however, the nodes are active in D-MIMO operation only when needed. These nodes further can streamline data processing, focusing solely on the physical layer for D-MIMO operation, bypassing the 5G RAN protocol stack processing (i.e., Radio Resource Control/Service Data Adaptation Protocol to Medium Access Control layers). The architecture also ensures integration with current cellular systems and reuse of existing deployments, thereby optimizing resource utilization. Our architecture holds the potential to be deployed in diverse environments such as stadiums, station squares, airports, railway stations, and large public gatherings to increase the capacity of cellular connectivity. 
The architecture also ensures integration with current cellular systems and reuse of existing deployments, thereby optimizing resource utilization.

This paper conducts a thorough analysis from an information-theoretic viewpoint to examine the potential capacity gains and their contributing factors. Our study demonstrates that despite this two-phase process, our architecture enhances system capacity, leveraging realistic 3GPP channel models for capacity analysis. The architecture we present is novel and to the best of our knowledge represents the first attempt to analyze the capacity of wireless mobile nodes participating in D-MIMO operation along with the BS for cellular connectivity, while also communicating with the BS through the access channel as users on the network. The paper is structured as follows: We start in Section~\ref{sec:architecture} by presenting an overview of the proposed architecture. Following that, we derive the capacity expressions of the proposed architecture for two phases in Section~\ref{sec:cap_analysis}. Finally, we analyze the theoretical capacity in Section~\ref{sec:simulation_insights} and conclude in Section~\ref{sec:conc}.

\section{System Architecture}\label{sec:architecture}

The proposed system comprises a BS surrounded by mobile wireless nodes participating in D-MIMO operation, forming a Mobile-Distributed Antenna Array (M-DAA) and serving a group of UEs within a cellular area. The operation of the proposed architecture is divided into two distinct phases. In Phase 1, depicted in Fig. \ref{fig:vtc_system_model}, a BS broadcasts data to the cooperating D-MIMO nodes, forming a downlink (DL) multi-user MIMO channel. Fig. \ref{fig:vtc_system_model} shows the handset UEs, and drones and vehicles with transceivers acting as mobile wireless nodes. Phase 2 communication occurs between the BS in cooperation with the D-MIMO nodes to the UEs, forming a D-MIMO channel, functioning as a multi-user MIMO Broadcast Channel (BC) or a single-user MIMO channel based on the transmission scheme used to serve the UEs. We assume that the nodes selected for D-MIMO operation are uniformly distributed in a circular arrangement with a radius $R$, with the BS at the center. In practice, the node selection or placement can be optimized based on ensuring good SNR links between the BS to nodes and nodes to UEs \cite{8683875,9615200}.

Phase 1 is the initial phase during which the D-MIMO nodes receive data from the BS, defining the capacity of the nodes participating in the M-DAA. All nodes in the M-DAA receive the same data. It is anticipated that this will be a relatively high-capacity link due to the nodes' proximity to the BS, with LOS or non-LOS conditions where path loss is fairly low. The key constraint here is that the nodes do not have the information to be transmitted in Phase 2 apriori and must obtain it from the BS. Once the information is received and decoded at each node, the BS and D-MIMO nodes encode, modulate, and precode the data based on the D-MIMO channel. Phase 1 follows a multi-user BC, except the BS transmits the same data to all nodes. In Phase 2, the nodes and BS collaborate to coherently joint transmit to the serving UEs. This collaboration is key in maximizing the system's capacity by providing improved beamforming gain. Since the data on each transmitter for a specific MIMO layer is the same, there is no interference between transmitters in the same layer, but inter-layer interference persists. The layer here is defined as the independent data stream sent either in Phase 1 or Phase 2. We consider that Phase 1 and Phase 2 share the same frequency band and each node and BS in the M-DAA possesses complete channel knowledge (CSI) between itself and the served UEs and serves only one UE at a time. There can be many possible architectures, but we choose this as our first exploration recognizing other architectures as potential future work.

\subsection{Challenges}

The two main challenges include synchronization and ensuring that CSI is available at the D-MIMO nodes and BS \cite{9492307}. Synchronization requires aligning both time references and carrier frequencies at the nodes and BS. In practice, BSs rely on the power-intensive Global Positioning System (GPS) for synchronization. However, this approach is untenable due to the battery-operated nature of D-MIMO nodes and the fact that the nodes might be in places where the GPS signal reception is weak. An alternative is wireless synchronization using a common beacon that transmits time and frequency synchronization signals among the M-DAA. This beacon can be either a separate node or the BS itself. Considerable theoretical and practical research \cite{7218555,6624252,8742232,3448623,9994246,8396683} has explored these techniques, achieving time accuracy in the picosecond range and frequency accuracy of less than 1 Hertz, approaching the precision of wired synchronization techniques. To ensure CSI is available at the nodes and BS, we can leverage the advantages of TDD channel reciprocity. This involves UEs transmitting reverse-link pilot signals, enabling the D-MIMO nodes and BS to perform channel estimation and compute the precoders for coherent transmission. However, the mobility of the UE impacts the promised gains, with higher mobility the channel changes rapidly and the precoding weights computed based on the outdated channel will not result in coherent transmission. Channel prediction algorithms \cite{9129558,8979256} would be needed depending on the mobility of UEs.

\section{Capacity Analysis}\label{sec:cap_analysis}
\subsection{Phase 1}
Considering $U$ nodes in the M-DAA, with $N_t^{BS}$ transmit antennas on at the BS and $N_r^{u}$ receive antennas at the node, the channel between the BS and the node is represented as $H_u$ with size $N_r^{u} \times N_t^{BS}$. The data at the BS is encoded and split into $N_s$ layers denoted as $S$, multiplied by the precoding matrix $F_u$ of size $N_t^{BS} \times N_s$ corresponding to each node and is transmitted by the BS. The received signal at the node $u$ is 
\normalsize
\begin{equation}
    Y_u = \sqrt{G_u} H_u X_u + \sqrt{G_u} \sum_{\substack{k=1,k \neq u}}^{U} H_u X_k + V_u, \label{eq1}
\end{equation}
\normalsize
where the transmitted symbol vector from the u-th D-MIMO node is $X_u= \sqrt{\frac{E_u}{N_t^{BS}}} F_u S$.
The $V_u \sim \mathcal{CN}(0, \sigma^2_{u} I_{N_r^{u}})$ represents the additive white Gaussian noise (AWGN) added at the receiver of the node. The entries of $H_u$ follow a complex Gaussian distribution $\mathcal{CN}(0,1)$ with their amplitude following Rayleigh fading. $G_u$ denotes the large-scale path-loss gain experienced by every entry of the channel. 
$E_u$ corresponds to the transmitted energy per symbol allocated per node and is subject to a sum constraint, $\sum_{u=1}^{U} E_u = E_s$. Expanding \eqref{eq1}, we get
\normalsize
\begin{equation}
Y_u = \sqrt{\frac{G_u E_u}{N_t^{BS}}} H_u F_u S + \sqrt{G_u} \sum_{\substack{k=1 \\ k \neq u}}^{U} \sqrt{\frac{E_k}{N_t^{BS}}} H_u F_k S + V_u. \label{ph1_eq2}
\end{equation}
\normalsize

The maximum number of data layers possible is equal to $N_s =\min(N_t^{\text{BS}}, N_r^u)$. Since the links between the BS and nodes have high SNR, we assume no CSI at the BS and thus choose precoders as $ \begin{bmatrix}
I_{N_s \times N_s } \\ 
0_{\left(N_t^{BS}-N_s\right) \times N_s}
 \end{bmatrix}$. Further, the overhead due to the channel estimation and precoder computation at the BS is eliminated. The equation \eqref{ph1_eq2} thus can be written as  
\begin{align*}
    Y_u = & \sqrt{\frac{G_u E_s}{N_s}} H_u S  + V_u.
\end{align*}
The mutual information between the BS and the $u^{th}$ node is $ I(Y_u,S) = H(Y_u) - H(Y_u|S)$,
where $H(Y_u) = \log|e\pi\Sigma_{Y_u}|$ and $H(Y_u|S)=H(V_u) =\log|e\pi\sigma^2_{u}I_{N_r^{u}}|$ and the co-variance of $Y_u$, i.e, $\Sigma_{Y_u}$ is
\normalsize
\begin{align}
    \Sigma_{Y_u} = & E\{Y_u Y_u^H\} = \frac{G_u E_s}{N_s}    H_u  H_u^H + \sigma^2_{Y_u}I. \label{eq:cov_ph1}
\end{align}
\normalsize
Here, the $H(X)$ is defined as the entropy of X \cite{1146355} and $\sigma^2_u$ is the AWGN noise power at node $u$. Then, based on \eqref{eq:cov_ph1}, the capacity of individual nodes is 
\normalsize
\begin{align*}
    I(Y_u,S) = \log_2 \left|\frac{E_s G_u}{N_s\sigma^2_{Y_u}}    H_u  H_u^H + I_{N_r^{u}} \right|. 
\end{align*}
\normalsize
Since all the nodes must receive the same data, the capacity (in b/s/Hz) of Phase 1 with bandwidth $B_1$, will boil down to the minimum capacity across all the receiving nodes, i.e.,
\normalsize
\begin{align*}
    C_1 =  B_1 \times \min [I(Y_u,S) ],
\end{align*}
\normalsize
and the system's capacity depends on the channel $H_u$ between the BS and the worst-case node. If some nodes have low channel gain, those nodes will have less rate bottle-necking D-MIMO capacity.

\noindent \textbf{Note}: The determinant of a matrix, represented as $|\cdot|$, where the ($\cdot$) represents any arbitrary matrix, is a scalar value that can be computed from its elements.

\begin{figure}[htpb]
    \centering
    \includegraphics[width=0.7\columnwidth]{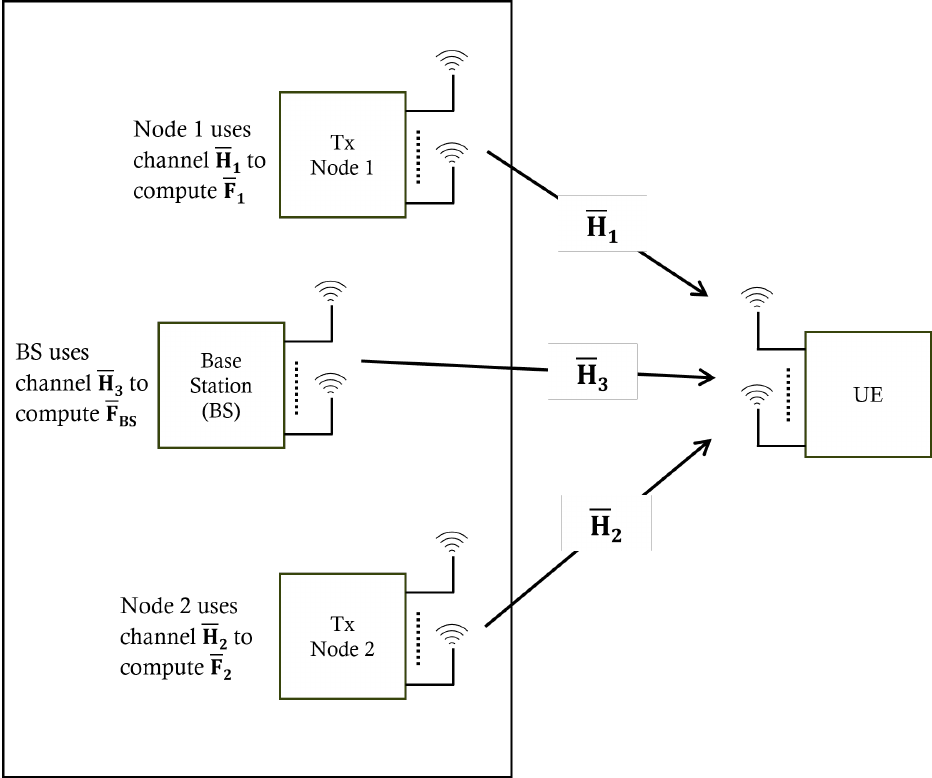}
    \caption{Phase 2 system showing CSI availability at nodes and BS for computing their ZF precoders}
    \label{fig:phase2}
    \vspace{-0.1 in}
\end{figure}
\subsection{Phase-2 (ZF)}
Given that each entity, i.e., each node and the BS of M-DAA knows the channel between itself and the serving UE, coherent transmission is enabled by agreeing upon a common transmission time. The channel estimation will be specific to each entity of the M-DAA, and the precoders will eliminate inter-layer interference caused by the channel between them, thereby providing the necessary beamforming gain at the receiver. The number of data layers at each entity of the M-DAA, \(\bar{N}_s\), in \(\bar{S}\) is determined by the minimum number of antennas of the entities of the M-DAA and serving UE, i.e., \(N_s = \min(N_t^{BS}, N_t^u, N_r^{UE})\). Since the data \(\bar{S}\) at each entity of the M-DAA is the same, each layer will have no interference from each transmitter. The channel between the BS and UE is represented by $\bar{H}_{BS}$ with size $N_r^{UE} \times N_t^{BS}$ and the node $u$ and UE is given by $\bar{H}_u$ with size $N_r^{UE} \times N_t^u$. $\bar{G}_u$ and $\bar{G}_{BS}$ are the large-scale path-loss gain experienced by every entry of the channel. Each node's precoder $\bar{F_u}$ and the BS's precoder $\bar{F}_{BS}$ will have sizes of $N_t^u \times \bar{N}_s$ and $N_t^{BS} \times \bar{N}_s$, respectively. Assuming the perfect CSI, the received signal at the UE is given by 
\normalsize
\begin{equation}
    \bar{Y}_{UE} = \sum_{u=1}^U \sqrt{\bar{G}_u} \bar{H}_u \bar{X}_u + \sqrt{\bar{G}_{BS}} \bar{H}_{BS} \bar{X}_{BS} + V_{UE}, \label{eq2_zf}
\end{equation}
\normalsize
where the transmitted symbol vector corresponding to that node and BS are given by 
\normalsize
\begin{align}
X_u  = \sqrt{\frac{E_u}{N_t^{u}}} \bar{F}_u \bar{S}, \ \& \
X_{BS}  = \sqrt{\frac{E_{BS}}{N_t^{BS}}} \bar{F}_{BS} \bar{S}. \label{eq3_zf}
\end{align}
\normalsize
Equation \eqref{eq2_zf} can be expanded using \eqref{eq3_zf} giving
\normalsize
\begin{align*}
    \bar{Y}_{UE}  = \sum_{u=1}^U \sqrt{\bar{G}_u} \bar{H}_u \sqrt{\frac{E_u}{N_t^{u}}} \bar{F}_u \bar{S}\  +   \sqrt{\bar{G}_{BS}} \bar{H}_{BS} \sqrt{\frac{E_{BS}}{N_t^{BS}}} \bar{F}_{BS} \bar{S} & \\  + \bar{V}_{UE}.    
\end{align*}
The mutual information between the M-DAA and the UE is
\begin{align*}
    I(\bar{Y}_{UE},\bar{S}) & = H(\Bar{Y}_{UE}) - H(\bar{Y}_{UE}|\bar{S}), \\
    & = \log_2|e\pi\Sigma_{\bar{Y}_{UE}}| - \log_2|e\pi\sigma^2_{UE}I_{N_r^{UE}}|,
\end{align*}
where the covariance matrix is $\Sigma_{\Bar{Y}_{UE}} = E\{\Bar{Y}_{UE} \Bar{Y}_{UE}^H\} =  \bar{H}\bar{F} \bar{F}^H \bar{H}^H + \sigma^2_{UE}I_{N_r^{UE}}$,
where, 
\begin{align*}
   & \Sigma_{\Bar{Y}} = E\{\Bar{Y}_{UE} \Bar{Y}_{UE}^H\} = \\
   &  \left(\sum_{u=1}^U\sqrt{\frac{\bar{G}_u \bar{E_u}}{N_t^{u}}}\bar{H}_u \bar{F}_u + \sqrt{\frac{\bar{G}_{BS} \bar{E}_{BS}}{N_t^{BS}}}\bar{H}_{BS} \bar{F}_{BS} \right) \times \\ &
   \left(\sum_{u=1}^U\sqrt{\frac{\bar{G}_u \bar{E_u}}{N_t^{u}}}\bar{H}_u\bar{F}_u + \sqrt{\frac{\bar{G}_{BS} \bar{E}_{BS}}{N_t^{BS}}}\bar{H}_{BS} \bar{F}_{BS} \right)^H+\sigma^2_{UE}I 
\end{align*}
Therefore, the capacity of the Phase 2 reduces to
\begin{align*}
      \log_2\bigg| \frac{1}{\sigma^2_{UE}}\left(\sum_{u=1}^U\sqrt{\frac{\bar{G}_u \bar{E_u}}{N_t^{u}}}\bar{H}_u\bar{F}_u + \sqrt{\frac{\bar{G}_{BS} \bar{E}_{BS}}{N_t^{BS}}}\bar{H}_{BS} \bar{F}_{BS} \right)  \\  \left(\sum_{u=1}^U\sqrt{\frac{\bar{G}_u \bar{E_u}}{N_t^{u}}}\bar{F}_u^H \bar{H}_u^H + \sqrt{\frac{\bar{G}_{BS} \bar{E}_{BS}}{N_t^{BS}}}\bar{F}_{BS}^H \bar{H}_{BS}^H  \right) + I\bigg| 
\end{align*} 
We opt for Zero-forcing (ZF) precoders $F_u$ to eliminate the inter-layer interference. These precoders diagonalize the argument of determinant maximizing the capacity \cite{Heath} and are given by pseudo-inverse of the entities' channels as 
\begin{align*}
    \bar{F}_{BS} = \bar{H}^{\dagger}_{BS} \ \& \ \bar{F}_u = \bar{H}^{\dagger}_u.
\end{align*}
The joint sum power constraints for the antennas on each node and BS are $\sum_{i=1}^{N_t^u} P_u^i \leq P_u$ and $\sum_{i=1}^{N_t^{BS}} P_{BS}^i \leq P_{BS}$. With proper normalization, the sum power constraints of nodes and BS are met, i.e., 
\normalsize
\begin{align}
E\{X_u X_u^H\} = \frac{E_u}{N_t^u} I_{N_t^{u}}  \ \& \  E\{X_{BS} X_{BS}^H\} = \frac{E_{BS}}{N_t^{BS}} I_{N_t^{BS}} 
\end{align} 
\normalsize
The capacity can be further reduced to
\begin{align*}
    \bar{N}_s \log_2\left( \frac{1}{\sigma^2_{UE}}\left(\sum_{u=1}^U\sqrt{\frac{\bar{G}_u \bar{E}_u}{N_t^{u}}} + \sqrt{ \frac{\bar{G}_{BS} \bar{E}_{BS}}{N_t^{BS}}} \right)^2 + 1 \right)
\end{align*}
This capacity expression indicates that each layer benefits from additional power with diversity, represented by the term \(\sum_{u=1}^U\sqrt{\frac{\bar{G}_u \bar{E}_u}{N_t^{u}}}\).

\begin{table}[h]
    \centering
    \begin{tabular}{|c|c|}
        \hline
        \textbf{Parameter} & \textbf{Value} \\
        \hline
        Pathloss model & UMi \\ 
        BS Tx power & 33 dBm \\   
        Node Tx power & 26 dBm \\ 
        BS height & 20 meters \\   
        UE height & 2 meters \\ 
        ($N^t_{BS}$,$N^t_{u}$,$N^r_{u}$,$N^r_{UE}$)  & (4,2,2,2) \\ 
        Bandwidth ($B_1, B_2$) & 10 Mhz \\ 
        Avg. height of buildings & 20 meters \\ \hline
    \end{tabular}
    \caption{Simulation Parameters}
    \label{tab:sim}
\end{table}
\vspace{-1em}
\section{Simulation Results and Insights} \label{sec:simulation_insights}
We adopt 3GPP channel models \cite{3gpp.38.901} for large-scale fading due to their inclusion of 3D environmental modelling. Urban Micro (UMi) is selected for its relevance to real-world settings like urban areas or station squares with NLOS settings in both phases. Path loss is determined based on antenna heights, carrier frequency, shadow fading, distances between BS and nodes in Phase 1, and distances between nodes and BS serving UEs in Phase 2.


First, we analyze the capacity of Phase 1 and Phase 2 individually with different numbers of nodes, i.e., 5, 10, and 20. Then, we combine the two phases and compare the D-MIMO capacity plots with the baseline (number of nodes = 0), where the BS directly communicates with the UE. 3GPP specifications \cite{3gpp.38.101.1,3gpp.38.101.2} define various power classes for UE (e.g., Vehicular, Handheld, High Power Non-Handheld, High-Speed Train Roof-Mounted UEs) that can be adopted for the node transmission power. All these users can be potential mobile nodes participating in D-MIMO operation. We considered the node transmission power of 26 dBm since this value was used for sidelink-UE and the base station transmission power of 33 dBm \cite{3gpp.25.942} defined in Table 5.1.8 of the 3GPP TS 25.942 specifications.


\begin{figure}[htpb]
    \centering
    \includegraphics[width=0.9\columnwidth]{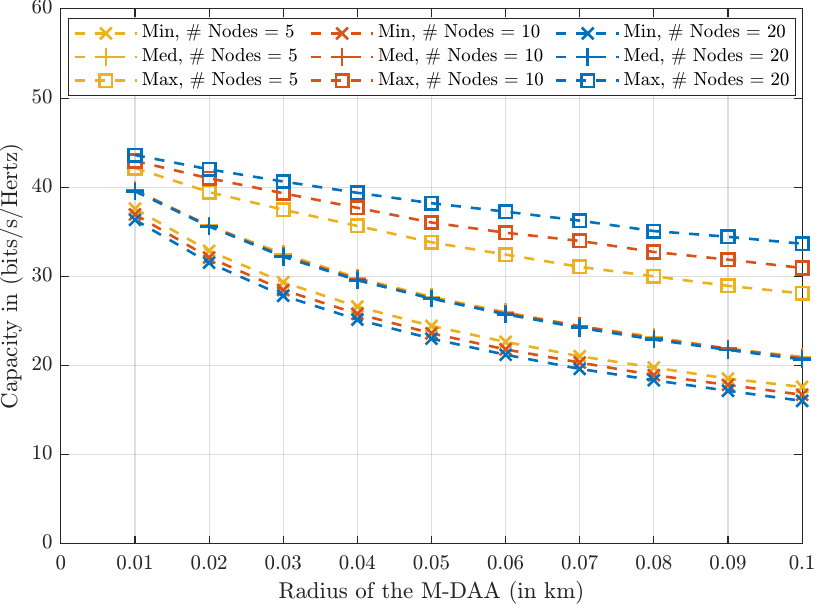}
    \caption{Average minimum, median, and maximum rates of nodes in Phase 1 under UMi settings}
    \label{fig:inter_squad_distance_nodes_phase1}
    \vspace{-0.2in}
\end{figure}

\subsection{Phase 1 Capacity Analysis and Insights}

Fig. \ref{fig:inter_squad_distance_nodes_phase1} illustrates the minimum (lowest rate), maximum (highest rate), and median (where half of the nodes exceed the median rate) average rates of all nodes selected by the BS for D-MIMO operation versus the radius of M-DAA. The nodes (drone, vehicular, or other handset UE) are defined with heights ranging from 2.5 to 25 meters, uniformly selected within this range. They are uniformly distributed in a circular arrangement with a radius \( R \), with the BS at the center. Node rates are higher when closer to the BS, influenced by the non-linear nature of the path loss gain versus distance equation \cite{3gpp.38.901}. Nodes farther from the BS experience greater path loss compared to closer nodes, leading to lower median and minimum rates. The plots indicate that as the number of nodes increases, the maximum rate rises due to an increased likelihood of nodes being nearer to the BS, while the minimum rate decreases because of a higher probability of nodes being farther from the BS. This implies that for Phase 1 to involve all nodes effectively in the D-MIMO phase, nodes should be uniformly distributed in a smaller geographic area within the circle (i.e., with a smaller radius) to ensure closer proximity to the BS and higher capacity for all nodes. In other words, the nodes should be chosen or placed around the BS to ensure uniformly good SNR links among nodes in Phase 1. 

The minimum rate node determines the capacity of Phase 1. Therefore, selecting a group of nodes with rates higher than the minimum rate node would increase the capacity of Phase 1. However, defining the minimum rate node as the capacity of Phase 1 allows all nodes to participate in the D-MIMO operation, enhancing the capacity of Phase 2. Selecting only a subset of nodes decreases the capacity of Phase 2.

\subsection{Phase 2 Capacity Analysis and Insights}
\begin{figure}[ht]
  \centering
  \begin{subfigure}{\columnwidth}
  \centering
    \includegraphics[width=0.9\linewidth]{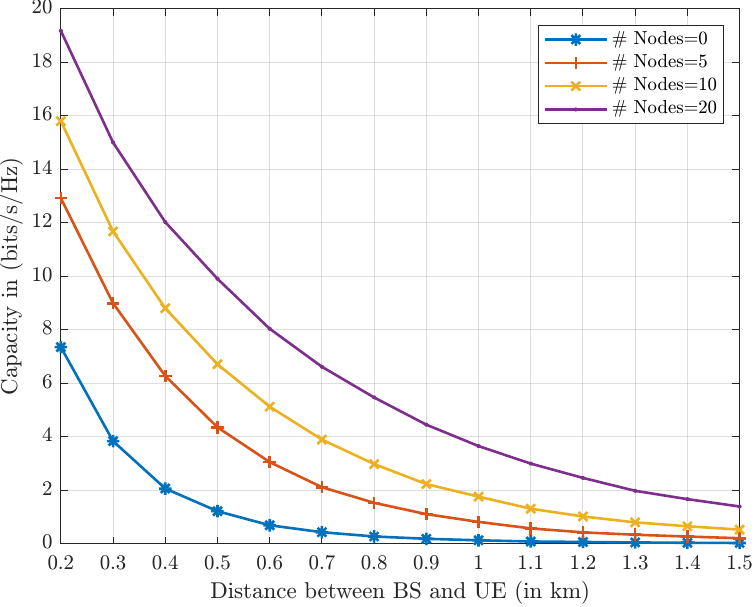}
    \caption{Phase 2 Capacity plots}
    \label{fig:inter_squad_dist_zf}
  \end{subfigure}
  \begin{subfigure}{\columnwidth}
  \centering
    \includegraphics[width=0.9\linewidth]{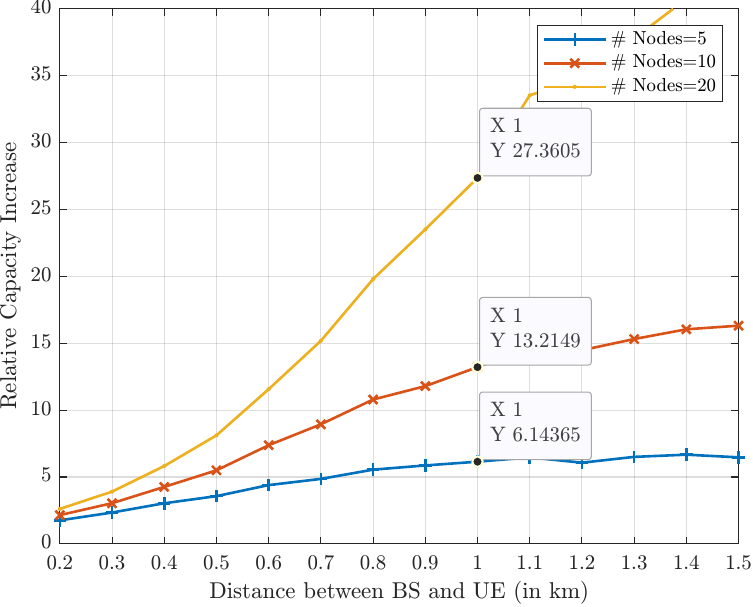}
    \caption{Phase 2 Capacity Relative increase over the baseline}
    \label{fig:range_improve_zf}
  \end{subfigure}

    \caption{Phase 2 average capacity and relative capacity increase over the baseline (with the number of nodes set to 0, ignoring Phase 1 cost) using ZF precoding under UMi settings.}

  \label{fig:overall_zf_cap_2x2}
  
\end{figure}

Fig. \ref{fig:overall_zf_cap_2x2} illustrates the average capacity of Phase 2 and the relative capacity increase compared to the baseline (when ignoring the cost of the Phase 1 transmission). The capacity improved by a factor of 6.14 with 5 nodes, 13.21 with 10 nodes, and 27.36 with 20 nodes at a 1 km distance between the BS and UE. Using ZF precoders, pseudo-inverses of the channel matrices, ensured that inter-layer interference was eliminated, allowing the received signal from all transmitters to be coherently summed. Each layer will have an additional power gain of \(\sum_{u=1}^U\sqrt{\frac{\bar{G}_u \bar{E}_u}{N_t^{u}}}\) over the baseline.  Additionally, as the distance between the M-DAA and UE increased, the relative capacity improvement over the baseline also increased. This also indicates that cell-edge users can significantly benefit from the proposed approach. To elaborate further, the 5 nodes contributed 10 antennas to the M-DAA. Additionally, each node provided an extra power of 26 dBm. This configuration facilitated the transmission of the same data from both the nodes and the BS to the UE. Similarly, in scenarios involving 10 and 20 nodes, there were 20 and 40 more antennas respectively compared to the baseline, along with the same 26 dBm additional power per node.

\begin{figure}[htpb]
    \centering
    \includegraphics[width=0.8\columnwidth]{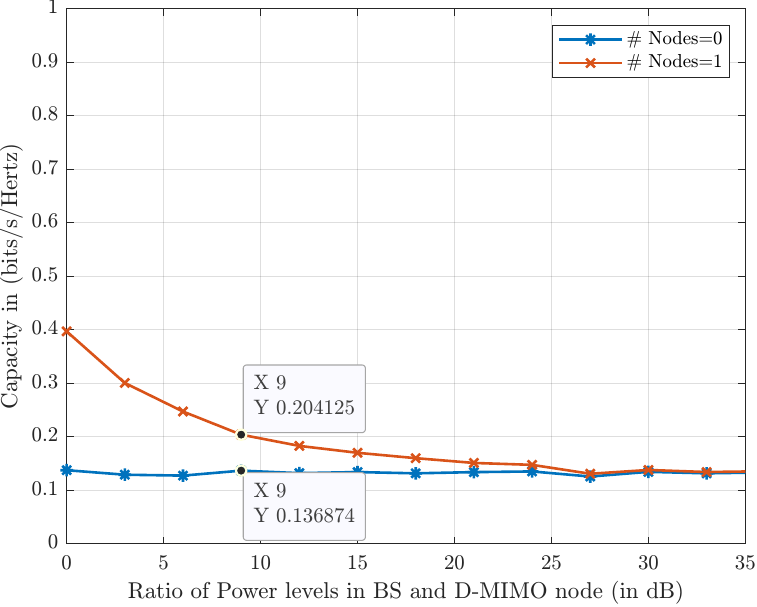}
    \caption{Phase 2 average capacity versus the D-MIMO node transmission power level}
    \label{fig:dmimo_node_tx_power_level}
    \vspace{-0.1in}
\end{figure}

We also explored the contribution of a node's capacity concerning its transmission power level compared to the BS in Phase 2. Fig. \ref{fig:dmimo_node_tx_power_level} shows the capacity with and without one node versus the node's transmission power level with UE placed at 1 km. The plot depicts that when the node's transmission power is below 9 dB compared to the BS power, the capacity contribution from the D-MIMO node is minimal. Thus, the node's transmission power plays a crucial role in defining the capacity of Phase 2.

\begin{figure}[htpb]
    \centering
    \includegraphics[width=\columnwidth]{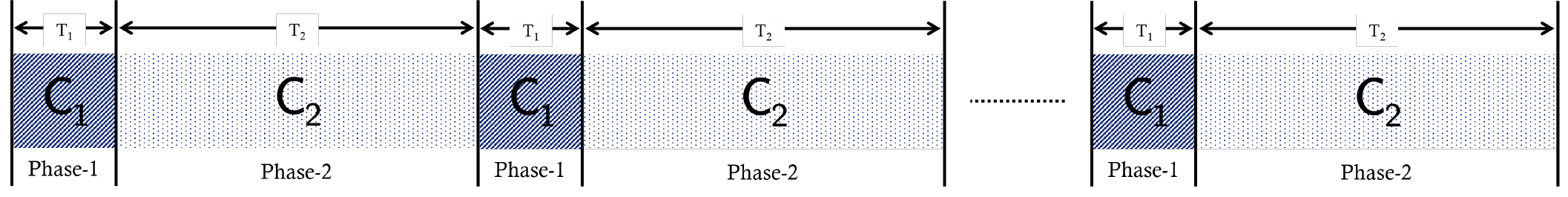}
    \caption{Operation of Phase 1 and Phase 2 through time}
    \label{fig:timing_sm_1}
    \vspace{-0.2in}
\end{figure}

\subsection{Phase 1 and Phase-2 Combined Capacity Insights}
Since D-MIMO operates in two phases, two time slots are required instead of one. Fig. \ref{fig:timing_sm_1} illustrates the timing diagram of the system operation. In this diagram, $T_1$ represents the time required for the BS to transmit its information bits to nodes at its maximum rate $C_1$ (i.e., the capacity of Phase 1). The time $T_2$ corresponds to the duration it takes for the nodes and BS to transmit all their information bits to the UE with their maximum rate $C_2$ (i.e., the capacity of Phase 2). After Phase 1 transmission, the same information bits $C_1 \times T_1$ (bits) are present at all the nodes and the BS. Now, in Phase 2 with its maximum rate $C_2$ (bits/s/Hz) through coherent transmission, to transfer the $C_1 \times T_1$ bits, the required time $T_2$ is
\begin{align*}
 T_2 = \frac{C_1 \times T_1}{C_2} .
\end{align*}
Therefore, $T_2$ is determined by extracting $C_1$ and $C_2$ from the simulations assuming $T_1=1$ second. Consequently, a fair comparison can now be made between the D-MIMO and the baseline. The baseline, augmented with time correction, will include an additional time $T_2$ for information transfer (i.e., baseline with its rate $C_B$ with time correction $C_B \times (1 + T_2)$) where $C_B = C_2$ when the number of nodes is equal to 0.

\begin{figure}
  \centering
  \begin{subfigure}{0.8\columnwidth}
  \centering
    \includegraphics[width=1\linewidth]{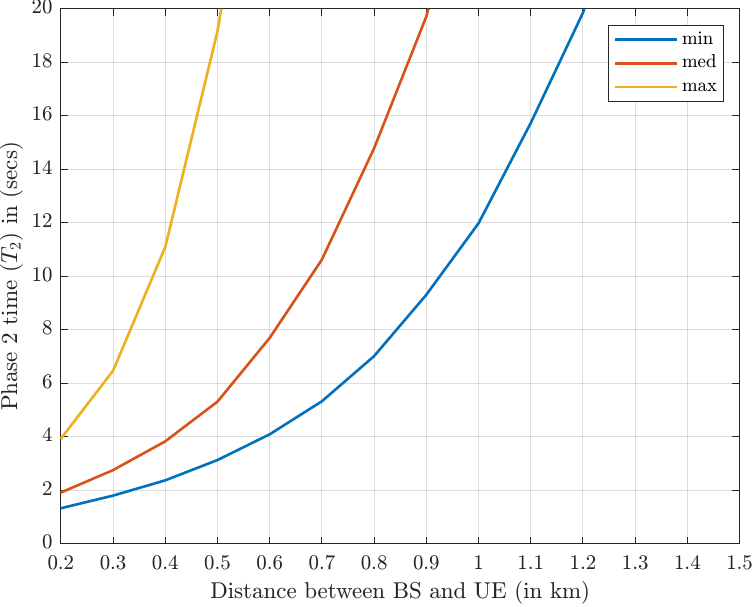}
    \caption{Phase 2 time $(T_2)$ to achieve same capacity as in Phase 1}
    \label{fig:overall_combined_cap_zf_time}
  \end{subfigure}
  \hfill
  \begin{subfigure}{0.9\columnwidth}
    \centering
    \includegraphics[width=\linewidth]{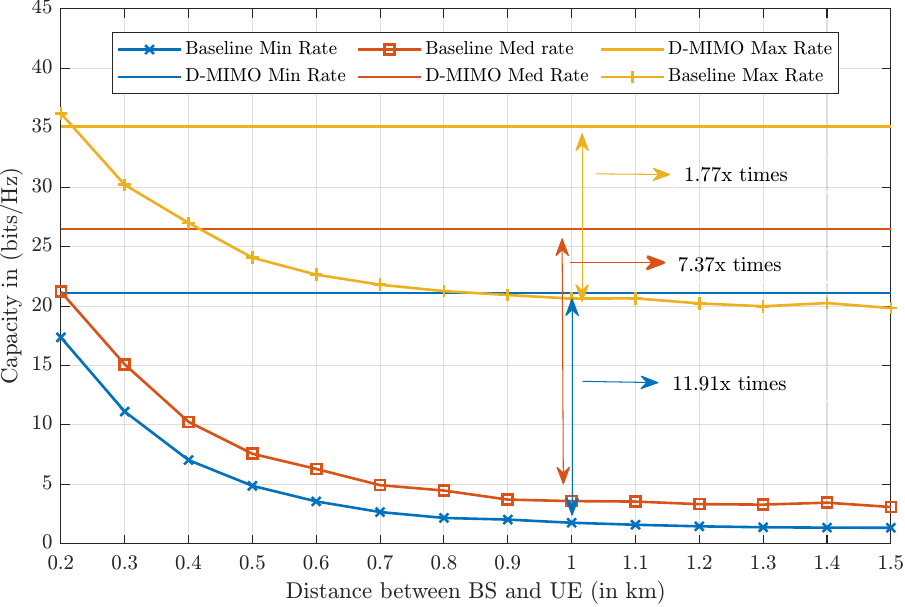}
    \caption{D-MIMO channel capacity comparison with the Baseline}
    \label{fig:overall_combined_cap_zf_cap}
  \end{subfigure}

  \caption{D-MIMO Capacity compared to baseline (with two-time slots) under UMi path-loss settings}
  \label{fig:overall_combined_cap_zf}
  
\end{figure}

Fig. \ref{fig:overall_combined_cap_zf_cap} illustrates the transfer of information bits (bits/Hz) from M-DAA to the UE in two time slots, one with a duration of $T_1 = 1$ second, and the other with a duration of $T_2$ seconds, as depicted in Fig. \ref{fig:overall_combined_cap_zf_time}. The flat curves represent the transfer of information bits in Phase 1 ($C_1 \times T_1$), which is the same as in Phase 2 ($C_2 \times T_2$), indicating the capacity of the D-MIMO system. The number of nodes considered is 10. Considering that the minimum rate defines the Phase 1 capacity, all nodes in the M-DAA operate during Phase 2. Consequently, the capacity of Phase 2 is high, requiring less time $T_2$ to transfer the information bits. However, when the median rate is used for Phase 1 capacity, only half of the nodes participate in Phase 2. In the scenario where the maximum rate is used for Phase 1, only the node with the maximum rate takes part in Phase 2. Hence, we observe a distinction in the duration of the D-MIMO channel, longer for the maximum rate, followed by the medium rate and then the minimum rate, as depicted in Fig. \ref{fig:overall_combined_cap_zf_time}. The largest gains over the baseline, achieved with time correction, are observed in the minimum-rate nodes, followed by the median-rate nodes. Conversely, the maximum-rate nodes exhibit the smallest gains. Specifically, the improvement values at a distance of 1 km are depicted in the plot, indicating enhancements of 11.91, 7.37, and 1.77 times, respectively. In terms of D-MIMO capacity, the maximum rate boasts the highest capacity, followed by medium and minimum rates. Thus, the plots indicate despite the two-phase operation, the capacity of D-MIMO is greater than the baseline with time correction. The plots also suggest that having at least one node with two-phase operation improves the total capacity of the system. Additionally, the improvements over the baseline increase as the distance between the BS and UE increases.

\section{Conclusion and Future Works} \label{sec:conc}

The capacity analysis of the proposed wireless D-MIMO architecture demonstrates a significant increase in system capacity based on realistic 3GPP channel models, even with the two-phase wireless operation without requiring changes in the network layout architecture. The results also indicate a greater relative capacity increase for UEs located farther from the BS compared to the baseline. Additionally, the proposed architecture ensures compatibility with current cellular systems and enables the reuse of existing infrastructure, thereby optimizing resource utilization. Future work of interest includes serving multiple users by scheduling through single-user MIMO or multi-user MIMO operations. Further, different architectures involving coherent and non-coherent joint transmission schemes including error rate analysis and system design with protocols and various realistic channel models such as 3GPP and air-to-ground (A2G), need to be explored. Another area of interest includes optimizing the placement or selection of nodes around the BS based on the surrounding environment to maintain good SNR links at all nodes.
\section{Acknowledgement}
Efforts sponsored by the U.S. Government under the Training and Readiness Accelerator II (TReX II), OTA. The U.S. Government is authorized to reproduce and distribute reprints for Governmental purposes notwithstanding any copyright notation thereon.

The views and conclusions contained herein are those of the authors and should not be interpreted as necessarily representing the official policies or endorsements, either expressed or implied, of the U.S. Government.




\bibliographystyle{IEEEtran}
\bibliography{reference}

\end{document}